# *Operando* Characterization and Molecular Simulations Reveal the Growth Kinetics of Graphene on Liquid Copper during Chemical Vapor Deposition

*Valentina Rein[1,*,‡], Hao Gao[2,‡], Hendrik H. Heenen[2,*,‡], Wissal Sghaier[3], Anastasios C. Manikas[4], Christos Tsakonas[4], Mehdi Saedi[5,6], Johannes T. Margraf[2], Costas Galiotis[4], Gilles Renaud[3], Oleg V. Konovalov[1], Irene M. N. Groot[5], Karsten Reuter[2], Maciej Jankowski[1]*

1 – ESRF – The European Synchrotron, 71 Avenue des Martyrs, 38043 Grenoble, France

2 – Fritz-Haber-Institut der Max-Planck-Gesellschaft, Faradayweg 4–6, 14195 Berlin, Germany

3 – University of Grenoble Alpes and CEA, IRIG/ MEM/NRS, 38000 Grenoble, France

4 – FORTH/ICE-HT and Department of Chemical Engineering, University of Patras, 26504 Patras, Greece

5 – Leiden Institute of Chemistry, Leiden University, P.O. Box 9502, 2300 RA Leiden, The Netherlands

6 – Physics Department, Shahid Beheshti University, Evin, Tehran, 1983969411, Iran

ABSTRACT: In recent years, liquid metal catalysts have emerged as a compelling choice for the controllable, large-scale, and high-quality synthesis of two-dimensional materials. At present, there is little mechanistic understanding of the intricate catalytic process, though, of its governing factors or what renders it superior to growth at the corresponding solid catalysts. Here, we report on a combined experimental and computational study of the kinetics of graphene growth during chemical vapor deposition on a liquid copper catalyst. By monitoring the growing graphene flakes in real time using *in situ* radiation-mode optical microscopy, we explore the growth morphology and kinetics over a wide range of $CH_4$-to-$H_2$ pressure ratios and deposition temperatures. Constant growth rates of the flakes' radius indicate a growth mode limited by precursor attachment, whereas methane-flux-dependent flake shapes point to limited precursor availability. Large-scale free energy simulations enabled by an efficient machine-learning moment tensor potential trained to density-functional theory data provide quantitative barriers for key atomic-scale growth processes. The wealth of experimental and theoretical data can be consistently combined into a microkinetic model that reveals mixed growth kinetics that, in contrast to the situation at solid Cu, is partly controlled by precursor attachment alongside precursor availability. Key mechanistic aspects that directly point toward the improved graphene quality are a largely suppressed carbon dimer attachment due to the facile incorporation of this precursor species into the liquid surface and a low-barrier ring-opening process that self-heals 5-membered rings resulting from remaining dimer attachments.

INTRODUCTION

Due to its outstanding electronic, optical, mechanical, and chemical properties, graphene is envisioned to catalyze the development of a next-generation array of products and devices in a wide range of applications.[1,2] Since its isolation in 2004,[3] research on and implementation of graphene has, in fact, already led to significant advancements in the electronics, medicine, sensor, energy, and space industries.[4,5] Chemical vapor deposition (CVD) is the state-of-the-art graphene production method.[6–8] In the graphene CVD process, a metal substrate surface, such as Cu, Ni, Pt, Fe, Ir, *etc*., acts as a catalyst for the decomposition of hydrocarbon precursor gas.[9] However, since the standard CVD approach to graphene growth is based on the use of a solid catalyst substrate, it suffers from multiple limitations. These solid substrates are often polycrystalline and display many defects and grain boundaries, which induce non-uniform and uncontrollable graphene nucleation and translate imperfections into the grown layer, severely undermining its quality.

As a response to the aforementioned challenges, liquid metal catalysts have been extensively explored since their introduction in 2012.[10] As shown in multiple studies and reviews,[11–14] CVD on a liquid substrate has a high potential for the advanced development of fast-growing, large-scale, single-crystalline graphene production with a reduced density of defects. The atomically smooth and homogeneous substrate surface is void of crystalline anisotropy and, therefore, prevents epitaxial influence on graphene flakes, as well as promotes a reduced, uniform, and controllable nucleation density, a fast mass transfer of surface carbon species and thus faster growth rates, and the self-assembly of graphene flakes.

The relatively weak adhesion of graphene to a molten surface is advantageous for the development of direct transfer technologies.[15–17] This would help to avoid a solidification step that is still present in the standard transfer procedure, which induces wrinkle formation and partially undermines the advantages of liquid substrates. We note, however, that the idea of liquid-based 2D material transfer is still in its infancy, and its realization on an industrial scale requires a significant amount of scientific advancement and technological innovation.

Among different metals, copper has been the most common and explored substrate for the graphene CVD process.[18–22] The main advantages are the low solubility of carbon atoms in Cu and their low diffusion barrier on Cu, facilitating the growth of the highest-quality large-area single-layer graphene (up to meter size)[23]. Due to its wide application, we have chosen Cu as a model liquid metal catalyst. However, it is worth mentioning that in the last years, many other liquid catalysts, such as Ag,[24] Cu-Sn,[25] Cu-Zn,[26] Cu-Ga,[27] *etc.*, have been shown as promising alternatives. These substrates benefit from relatively low melting point temperatures and other intriguing properties, such as a low binding force which minimizes wrinkle formation in the case of liquid Ag.

The elementary processes that occur during graphene's CVD growth on solid or liquid metal catalysts, such as copper, are schematically illustrated in Figure 1 and explained in detail in the Supporting Information (SI). While the parameters (*e.g.*, pre-exponential factors and activation energies) for these processes are relatively well established for solid substrates,[28–31] very little is known for liquid substrates, and the values of, *e.g.*, surface diffusion of the different species, are expected to differ by orders of magnitude from those on solid surfaces. Due to the high complexity of the growth mechanism, the actual optimization of growth parameters on liquid catalysts is still quite challenging, especially as the detailed growth mechanism and its differences from the one on the established solid catalyst substrates are not well known.

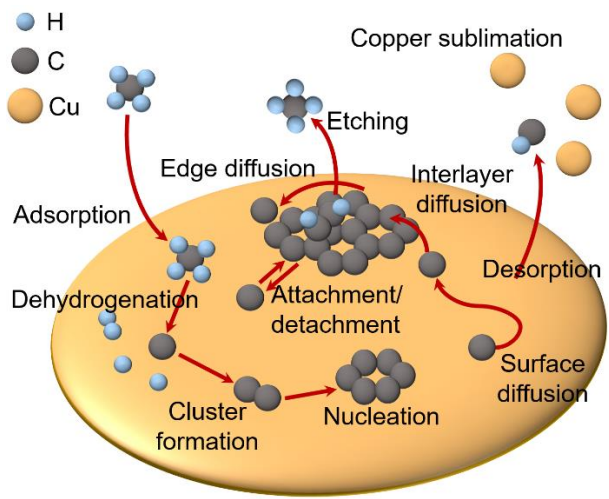

**Figure 1.** General illustration of the graphene CVD growth process on solid or liquid Cu. The detailed description is provided in the SI.

Until recently, studies on graphene grown on liquid metals were primarily restricted to *ex-situ* post-growth characterization that entails a significant loss of information.[10,32] This limitation was due to the harsh experimental conditions (*e.g.,* high evaporation rates of molten metals, high pressure, reactive gas environment, substrate temperature around 1400 K), where the application of standard ultra-high vacuum, electron-based techniques was challenging. Such limited experiment conditions hindered the extraction of quantitative information, *e.g.*, activation energies. Technological development in the last decade has enabled many characterization techniques to be applied *in situ*.[24,33] However, due to the lack of sensitivity and/or the limitations in realizing relevant growth conditions, accurate analysis of the growth kinetics was still problematic.

Implementing radiation-mode optical microscopy for *operando* and *in situ* investigations can be considered a significant advancement in that regard. Using infrared and visible light, which is not significantly absorbed by gases, has enabled direct observation of graphene growth in real-time.[34] This approach visualizes the growth of single-layer graphene flakes by exploiting the difference in emissivity between graphene and liquid copper at high temperatures (~1370 K). Moreover, this method applies to studies on liquid copper, where the movement of graphene flakes on the liquid surface and the high evaporation rate of liquid metal brings additional complexity.[14,35] To take advantage of these benefits, a CVD setup and optical system were designed to sensitively manipulate the experimental conditions and follow the graphene flakes' motion and growth kinetics in detail.[36] The latter gives access to a deep mechanistic understanding that aids in controlling the growth parameters and optimizing the synthesis of large-area single-crystalline graphene domains which has been lacking so far.

In our combined experimental and computational study, we rigorously assess the growth mechanism and kinetics of graphene domains on liquid Cu. On the experimental side, we employ the aforementioned CVD reactor designed for *in situ* radiation-mode optical microscopy to study a wide range of growth conditions. On the computational side, the use of machine-learning potentials as fast surrogates to first-principles calculations enables a reliable sampling of the liquid state. Otherwise intractable at the first-principle level, these large-scale simulations give access to quantitative free energy barriers for various key growth processes. Matching the experimental and computed data within a microkinetic model, we arrive at a mixed growth mechanism that is partially governed by both precursor availability and precursor attachment. The most crucial difference to growth on solid Cu seems to be the facile incorporation of carbon dimers into the liquid substrate, the consequences of which may also rationalize the improved graphene quality.

RESULTS AND DISCUSSION

**Procedure and quality control.** Graphene is grown in a customized CVD reactor[36] on molten copper at a total pressure of 200 mbar using methane as the precursor gas in an Ar/H$_2$ atmosphere (see the Methods section for further experimental details). The effect of the absolute H$_2$ pressure was checked (see the SI, Figure S1), and the default H$_2$ partial pressure used ensures optimum growth conditions. Consequently, in the rest of the paper, the partial pressure ratio $p_{CH4}/p_{H2}$ is only varied by varying the partial pressure of $p_{CH4}$ at constant default $p_{H2}$. The growth procedure is illustrated in Figure 2 and Movie S1 in the SI. We first apply a high partial pressure of methane ($p_{CH4}/p_{H2}$ between 1.81−2.72×10$^{-2}$, Figure 2a) to facilitate nucleation and accelerate the growth of the first flakes. After following the evolution of the flakes for a few minutes until their coalescence, the methane flow is turned off to initiate etching of the flakes in the H$_2$/Ar atmosphere ($p_{CH4} = 0$, Figure 2b). As soon as only a few tiny islands are left on the surface, the methane flow is changed to an intermediate partial pressure value (*e.g.*, $p_{CH4}/p_{H2} = 1.27 \times 10^{-2}$, Figure 2c and d), and the growth process is carefully followed and analyzed. Note that in the regime of medium flows ($0.54 < p_{CH4}/p_{H2} < 1.81 \times 10^{-2}$), continuous nucleation still occurs, although its density and rate are reduced. In order to cover a broad growth rate range, the cycle of etching and regrowth at different $p_{CH4}/p_{H2}$ was repeated several times for five temperatures $T = 1368, 1399, 1416, 1433,$ and $1456$ K within the instrumentally accessible range.

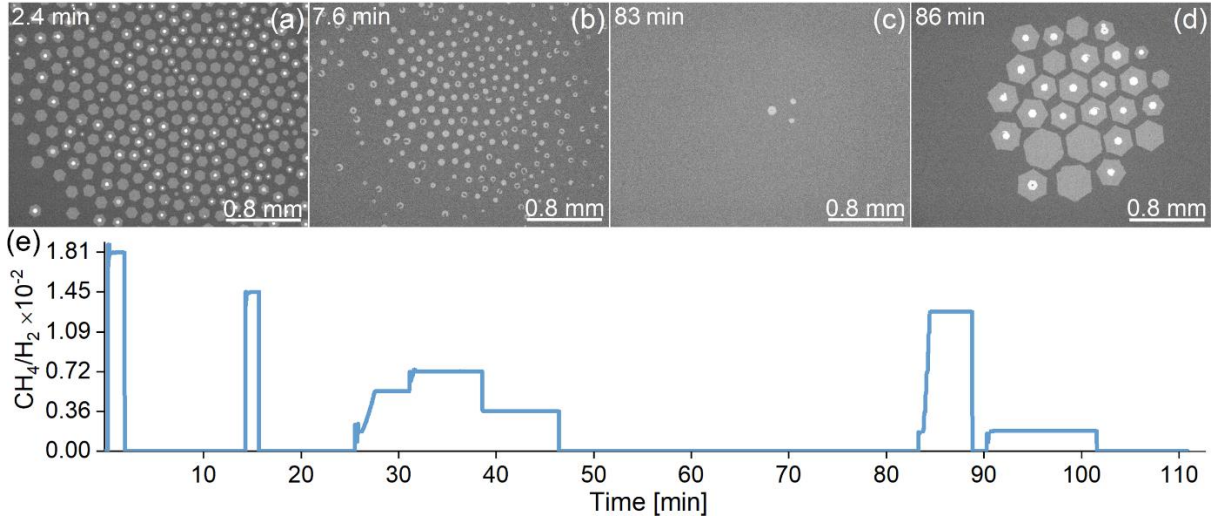

**Figure 2.** *Top:* Experimental steps of CVD graphene growth on liquid Cu: (a) initial nucleation and growth of flakes at a high partial CH$_4$ pressure ($p_{CH4}/p_{H2}$ between $1.81-2.72 \times 10^{-2}$); (b) etching ($p_{CH4} = 0$); (c) and (d) regrowth with a lower flow of methane (here, $p_{CH4}/p_{H2} = 1.27 \times 10^{-2}$). The time counts from the moment the methane valve is opened initially (before image (a)). See also Movie S1. *Bottom:* (e) Time evolution of the gas pressure ratio corresponding to images (a)-(d).

For each image frame, the averaged flake area $A$, the diameter or long diagonal (for irregular shapes), the circumference $L$, and the circularity $(4\pi A/L^2) \times (1-0.5/(L/2\pi+0.5))^2$ of the flakes are extracted using the MATLAB image processing toolbox. Quality control of the grown graphene samples is performed by *ex-situ* Raman spectroscopy after solidification and standard wet transfer onto Si/SiO$_2$ wafers. Due to this procedure, the final morphology is undulated as it replicates that of the solidified copper. The Raman spectra confirm the growth of single-layer graphene through a ratio of intensities of two characteristic peaks $I_{2D}/I_G$. The corresponding analysis is provided in the SI, Figures S2−S4. The detailed Raman characterization of the graphene obtained in the reported setup, including mapping of an entire flake (~400 μm), has been shown in a previous publication.[14] There, we found that the density of structural defects within flakes is very low under common growth conditions. Due to the atomically flat liquid surface, the graphene flakes do not take over structural defects from an otherwise polycrystalline substrate, which is the case for solid Cu. We do not expect such a low defect density to significantly alter our results.

**Flake morphology.** First, we visually examine the variation of the morphology of growing flakes and find it to be dependent on the growth time (which determines the flake size) and partial pressure of the precursor. Similar observations have been reported by different experimental and theoretical (phase-field modeling) studies.[32,37–40] The observed morphological behavior can be roughly categorized into five modes depending on the ratio between methane and hydrogen pressures $p_{CH4}/p_{H2}$ (Figure 3). A quantitative illustration of the shape evolution with the flake size for different pressure ranges can be found in the SI (Figures S5 and S6). We note that we do not see any prominent impact of temperature on the morphology within the ~100-degree range accessible with our instrument but rather on the growth and etching rates, as will be shown in the following subsection.

At the highest CH$_4$ flows ($p_{CH4}/p_{H2}$ = 1.81–2.72×10$^{-2}$, where spontaneous nucleation occurs, Figure 3a, b), flakes maintain a well-defined circular shape without noticeable changes during growth. When the content of CH$_4$ is lower but still relatively high ($p_{CH4}/p_{H2}$ = 1.45–1.81×10$^{-2}$, Figure 3c, d), flakes initially grow as perfect hexagons and later develop slightly concave edges (after 5 minutes). For medium CH$_4$ flow ($p_{CH4}/p_{H2}$ = 0.73–1.45×10$^{-2}$, Figure 3e, f), the transition from the initial hexagonal shape to a concave dodecagon is faster, with the external angle reaching 10° (Figure S5b). At low CH$_4$ flow ($p_{CH4}/p_{H2}$ = 0.18–0.54×10$^{-2}$, Figure 3g, h), C species flux is insufficient for nucleation, but existing graphene flakes continue to grow, forming sharp concave dodecagons with external angles of up to 20° (Figure S5a). In parallel, the flakes start to etch at their centers where the availability of C species is minimal. Various structural defects might also initiate etching.[41] When methane flow is turned off ($p_{CH4}$ = 0, Figure 3i, j), etching begins at the outer edges and in the middle of the flakes, targeting defects (based on visual analysis). In this pure etching regime, the reverse transition from dodecagon to hexagon and then to circle is observed.

The processes governing the flake shape are generally attributed to concentration gradients of surface carbon species and their diffusion along the flake edge.[38–42] At high methane pressure, a homogeneous distribution of carbon species on the liquid Cu catalyst leads to an isotropic circular growth.[11,43] However, zigzag edges are energetically more favored than armchair ones, and over time, edge diffusion drives flakes toward their thermodynamic equilibrium hexagonal shape.[41–45] As hexagonal shapes develop, corners of the hexagons begin to benefit from higher precursor concentration, resulting in protruded corners that form a dodecagon shape at the later growth stages.[38] Edge diffusion, while still favoring hexagons, becomes limited as flakes grow in size, resulting in less compact shapes.[46] These effects are sensitive to reactant concentrations, and the shape transitions are therefore commonly associated with transport limitations, which implies a mechanistic relevance of surface diffusion and/or CH$_4$ activation that both determine precursor availability.

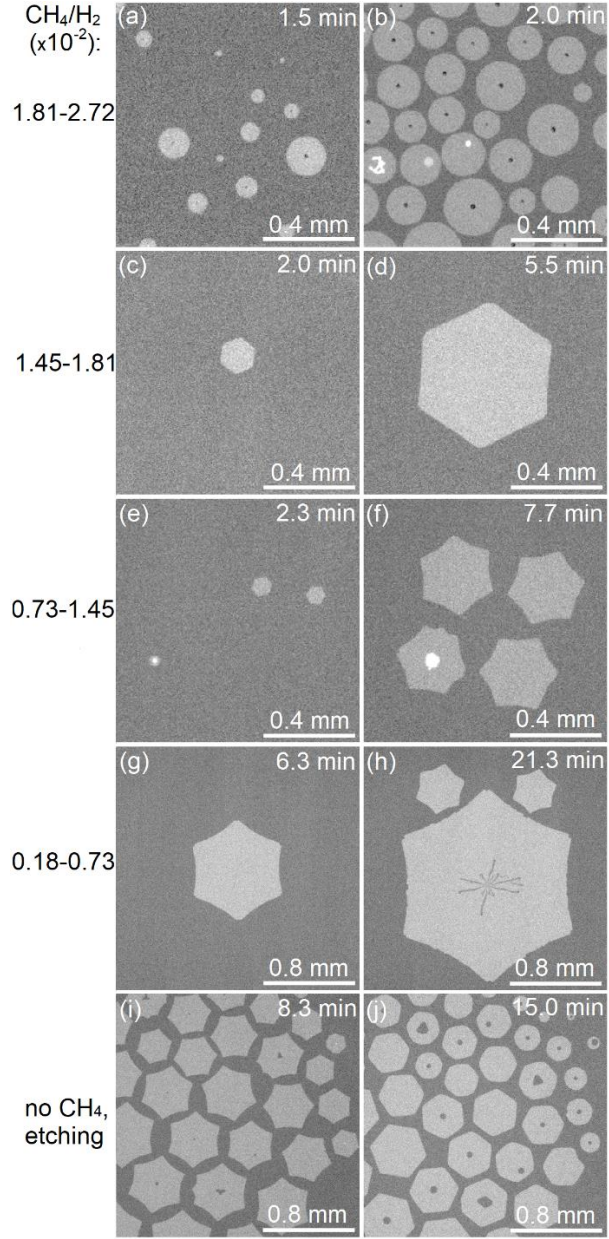

**Figure 3.** Exemplary radiation-mode optical microscopy images of the typical morphologies for different methane/hydrogen partial pressure ratios. The zero time is the moment the methane flow is set to the indicated value; left images are at earlier, right images at prolonged exposure times.

**Growth rates.** We define the flake growth (or etching) rate as the change in lateral flake size over time. Since the shape of the graphene flakes is not constant, we consider as a parameter of the lateral size the effective radius $R_{\text{eff}}$ described as the ratio between the flake area $A$ and circumference $L$,

$$R_{\text{eff}} = \frac{2A}{L}. \qquad (1)$$

As demonstrated in Figure S7, the average $R_{\text{eff}}$ is found to increase linearly with time, which means that the corresponding areal growth rates are size-dependent, as also shown in Figure S8. Surprisingly, the linear trend of $R_{eff}$ is traceable over broad pressure and temperature ranges without deviations and despite the shape transformations discussed above. Moreover, for the case of etching, a linear decrease of $R_{eff}$ is found, as seen from the negative slope of some curves in Figure S7 at a CH$_4$ flow with $p_{CH4}/p_{H2}$ below 0.18−0.36×10$^{-2}$. Note that we do not consider the optically inaccessible nucleation stage, but instead, only later growth stages that are at the same time still relatively far from the flakes' coalescence and closure of the layer so that most of the flakes have some degree of freedom, as illustrated by exemplary Movie S1. Indeed, a noticeable deviation of the lateral growth rates from the observed linear evolution of the radius as a function of time appears at these latest coalescence and closure stages, as demonstrated in Figure S9 and Movie S2.

The fact that $R_{eff}$ increases at a constant rate across a wide range of flake sizes (ranging from 15 μm up to 1.6 mm in diameter) suggests that growth takes place in an attachment-limited (also called reaction- or edge-kinetics-limited) regime. According to theoretical models for constant flake shapes,[47–49] the radial growth rates in this regime are proportional to both the extent of the bare Cu surface and the concentration of the reactant. Since we find equivalent growth rates of flakes with equivalent $R_{eff}$ but different shapes, there may be a cancelation between faster-growing areas and slower-growing areas in the case of the non-compact shapes so that the effective radius stays shape-independent. Nevertheless, the finding of a linear growth rate is a strong indicator for the mechanistic relevance of precursor attachment, which is thus at variance with the relevance of precursor availability derived from the analysis of the flake morphology changes with varying CH$_4$ flow.

**Apparent activation energies**. To investigate this conflicting situation, we next systematically study the variation of the linear growth rates as a function of the pressure ratio $p_{CH4}/p_{H2}$ and temperature $T$. As presented in Figure 4a, up to some critical value of $p_{CH4}/p_{H2} \approx 1.45–1.81 \times 10^{-2}$ (the value increases with $T$), the growth rates are found to increase almost linearly with $p_{CH4}/p_{H2}$ at all $T$. Above $p_{CH4}/p_{H2} = 1.63 \times 10^{-2}$, this evolution with pressure saturates towards lower rate values, whereas towards lower partial pressure ratios a zero growth rate is reached at $p_{CH4}/p_{H2} \approx 0.27 \times 10^{-2}$. At this point, the concentration of carbon species $C$ should correspond to the equilibrium concentration $C_{eq}$, and a balance between attachment and detachment rates is reached. The observed linearity of the growth rates above this pressure ratio can then be understood within classical film growth theory, which predicts the edge growth rate to be proportional to the degree of supersaturation ($C-C_{eq}$).[50] The deviation from linearity toward the highest partial pressure ratios finally arises both from the saturation of the Cu surface with C species and the dual role of H$_2$, as elaborated in the SI (Figure S1).

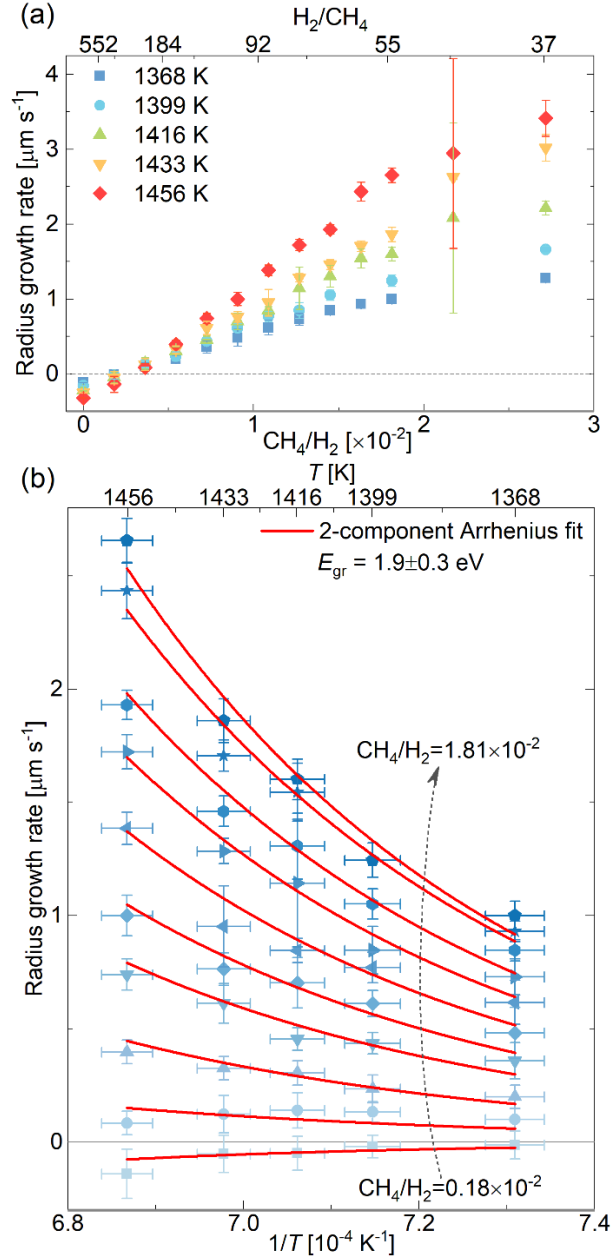

**Figure 4.** Growth rates of graphene flakes on liquid Cu: (a) lateral growth rates plotted as a function of partial pressures and $T$ for low $p_{CH4}/p_{H2}$ ratios (the larger error bar at $2.17\times10^{-2}$ results from a poor statistics for this point); (b) lateral growth rates as a function of $1/T$ for various $p_{CH4}/p_{H2}$ ratios $\leq 1.81\times10^{-2}$.

Analysis of the temperature dependence of growth rates provides complementary insight into rate-determining steps of the activated growth mechanism. Here, we focus on the most relevant partial pressure ratio regime leading to linear growth rates and show the corresponding Arrhenius plots of the growth rates in Figure 4b. As expected, the growth rate increases with the substrate temperature. However, as can be seen, the dependence is non-linear in the Arrhenius coordinates, which reflects a varying dominance of at least two rate-controlling steps over the range of partial pressure ratios probed. From the overall decrease of the growth rate with temperature toward the lower partial pressure ratios, we specifically assign this to increasing dominance of adversary etching, *i.e.,* the detachment of C species due to etching by hydrogen. We correspondingly fit the data with a two-component Arrhenius equation for growth (gr) and etching (et):[42]

$$\text{GR} = ap_{\text{CH4}}e^{\frac{-E_{\text{gr}}}{kT}} - bp_{\text{H2}}e^{\frac{-E_{\text{et}}}{kT}}, \qquad (2)$$

where GR is the growth rate, $a$ and $b$ are pre-exponential coefficients, $k = 8.63 \times 10^{-5}$ eV K$^{-1}$ atom$^{-1}$ is the Boltzmann constant, and $E_{\text{gr}}$ and $E_{\text{et}}$ are the apparent activation barriers for growth and etching, respectively. We specifically extract $E_{\text{et}}$ and constant $b$ from the 'pure etching' regime without CH$_4$ present (Figure S10), where the data indeed exhibits an essentially linear Arrhenius dependence, *cf.* Figure 4b. With the determined $b$ and $E_{\text{et}} = 2.0 \pm 0.1$ eV, we then fit the data points from the linear $p_{\text{CH4}}/p_{\text{H2}}$ range (between 0.18–1.81×10$^2$) in Figure 4b to Equation 2 to obtain $E_{\text{gr}} = 1.9 \pm 0.3$ eV. This apparent activation barrier for growth on the liquid Cu is slightly lower than the values of 2.3–2.6 eV that were previously estimated for solid copper, yet without considering an adversary etching process.[28,31]

**Free-energy simulations and microkinetic model.** In order to connect the derived apparent activation barriers to an elementary-process mechanism and resolve the conflicting insight into the relevance of precursor attachment (growth rate analysis) and precursor availability (flake morphology analysis), we now turn to computer simulations. Specifically, we employ an efficient machine-learning moment-tensor potential accurately trained to density-functional theory data (see Methods section). This potential enables extensive sampling, which is necessary to simulate the liquid Cu surface realistically and is unfeasible using density functional theory calculations directly. In the first step, we evaluate the hypothesis of reaction-limited growth with attachment processes as the rate-limiting step. Specifically, we conduct free-energy calculations at 1370 K of the attachment process of a monomer or dimer carbon species as typical precursors[20,51,52] to both dehydrogenated[53] zigzag and armchair graphene edges. The simulation of these idealized edges ignores the possible influence of defects or imperfections as well as a simultaneously occurring dehydrogenation during attachment (see also discussion in the SI). However, due to the creation of many dangling bonds, we assume this step in the flake growth to be the least favorable and, thus, most limiting. The corresponding free energy profiles for the zigzag edge are shown in Figure 5a,b (see the SI and the Methods section for further details), revealing attachment and detachment barriers of 1.51 and 1.87 eV for the monomer and 1.38 and 1.99 eV for the dimer, respectively. Essentially, identical values and free energy profiles are obtained for the armchair edge (Figures S14, S15, and Table S1). This equivalency of the two flake edges has also been observed on solid Cu[54] and excludes a possible influence on the growth rate by the less stable armchair edge,[55,56] which may become more prominent with changing flake shape or growth regime.[57,58]

The computed detachment barriers of 1.87 and 1.99 eV for monomer and dimer agree very well with the experimentally deduced apparent activation barrier for etching (2.0±0.1 eV, see above), which suggests carbon detachment as a solely rate-limiting mechanistic step. In contrast, the simulated monomer or dimer attachment barriers are 1.51 and 1.38 eV, respectively, somewhat smaller than the experimental apparent activation barrier of 1.9±0.3 eV for growth (see above). This slight discrepancy indicates that the growth kinetics might not be entirely controlled by precursor attachment, exactly as also deduced from the analysis of the flake morphology changes.

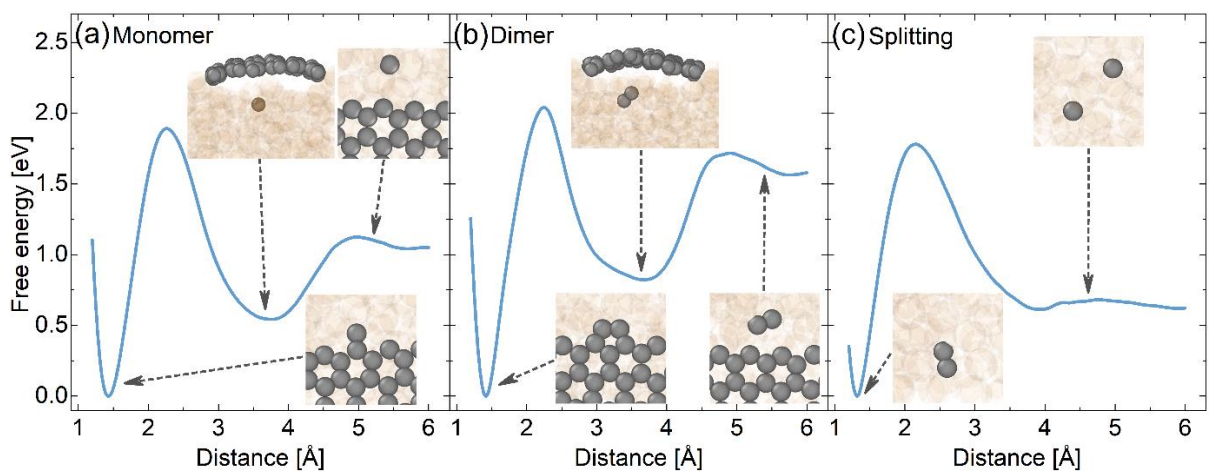

**Figure 5.** Free energy profiles of the attachment/detachment of (a) a carbon monomer and (b) a dimer to/from graphene zigzag edges, and (c) dimer dissociation and monomer association from the umbrella sampling simulations conducted at 1370 K. Representative configurations are shown as insets, where the carbon and copper atoms are colored gray and transparent-orange, respectively. Most insets are top views, except for two side views of structures showing the local minima on the free energy surfaces of monomer/dimer attachment/detachment, characterized by the precursor location under the graphene flake. Note that free energy differences stated in the text and used in the microkinetic model are based on the integration of reactant and product basins, as elaborated in the SI.

Turning our attention, therefore, to precursor availability, a first intriguing aspect can already be discerned from the attachment/detachment free energy profiles of the monomer and dimer shown in Figure 5a–b. In both cases, there is a pronounced local minimum structure in which the precursor is stabilized within the liquid Cu and below the graphene sheet (see validation and details of the minimum structure in the SI). Attachment will, therefore, unlikely proceed from a freely diffusing state but instead out of this subsurface state for both monomer and dimer. Following these similarities in the attachment mechanism, the attachment barriers are also very similar for both reactants (Table S1, S3, and Figure S15). This is in stark contrast to the situation for solid Cu, where sub-surface configurations for the dimer are prohibitively unstable, and a robust stabilization underneath the graphene flake is only found for the monomer.[52]

Consequently, the attachment barrier for the monomer is ~0.5–0.7 eV higher than for the dimer, and graphene flake growth at solid Cu proceeds predominantly through dimer attachment.[52,57]

With the similar monomer and dimer attachment barriers at liquid Cu, it is, therefore, rather the steady-state populations of the two species that determine the growth mechanism. These populations, *i.e.*, their availabilities, result not only from the balance between depletion due to flake attachment and C monomer formation due to dissociative methane adsorption but also from the continuous interconversion of the two species by monomer association and reverse dimer dissociation processes. As shown in Figure 5c (and S16 and S17), we compute the carbon dimer state to be only moderately more favorable by a free-energy difference of ~0.3 eV (as compared to ~0.8 eV at solid Cu(111)[53] and the free energy barrier for dimer formation to be as high as 1.44 eV.

If we combine these numbers with the experimental parameters for temperature and pressure within a simple mean-field microkinetic model to assess the contribution of precursor availability to the overall growth kinetics (see SI for details and a critical discussion), we obtain complete agreement with the measured apparent activation barrier for growth $E_{gr}$ when we assume high barriers for methane dissociation in the range 1.5–2.2 eV. This range is fully compatible with previous estimates on solid Cu,[31] and in this range, we then indeed find the kinetics to be only partially governed by precursor attachment (<25%, according to a degree-of-rate-control analysis[59]). This partial attachment rate control rationalizes the experimentally observed flake size-independent $R_{eff}$ growth rates. At the same time, the additional partial limitation of precursor availability due to the high methane dissociation barrier leads to non-saturated precursor coverages that can account for the build-up of local concentration gradients around the graphene flakes that lead to the observed range of $p_{CH4}/p_{H2}$-dependent flake morphologies (Figure 3).

The contribution of dimer attachment to the graphene flake growth predicted by the microkinetic model is only of the order of 10% (Figure S23) and thus dramatically lower than on solid Cu. Since each dimer attachment will initially lead to the formation of a defect motive in the form of a 5-membered ring (see Figure 5), this lowered contribution could already rationalize the improved graphene quality obtained at liquid Cu catalysts. Moreover, we find that the liquid Cu surface facilitates a ring-opening process with a barrier of 1.35 eV (SI Figures S18 and S19, as well as Tables S2 and S3) that is thus lower than the one of the actual dimer attachment. This process makes the formation of a 5-membered ring reversible and acts as a defect-healing mechanism, confirming a previous hypothesis derived from observations in *ab initio* molecular dynamics simulations.[60]

CONCLUSIONS

We investigated the CVD growth of graphene domains on a liquid copper catalyst by using real-time *in situ* optical microscopy in radiation mode in combination with free-energy simulations and a microkinetic model. We found that the flake morphology (varying between hexagonal and circular shapes) is almost independent of the temperature (in the range $T$ = 1368–1456 K) but depends strongly on the methane pressure and flake size. At the same time, the lateral growth rates at constant pressures and temperatures reveal no time or size dependence. Both types of finding cannot be reconciled with a simple growth process controlled only by precursor availability as featured on solid Cu.[31,61]

Detailed Arrhenius analysis of the experimental data demonstrates that, first of all, the competing process of detachment/etching with an apparent activation barrier of 2.0±0.1 eV must be considered to understand the overall growth kinetics. In addition, extensive first-principal-quality free energy simulations indicate that both the attachment of carbon-active species and methane activation contribute to the measured apparent activation energy of 1.9±0.3 eV for growth. Significant differences in the detailed attachment process provide thereby first leads to understanding the improved graphene quality compared to solid Cu catalysts. Due to the facile incorporation of both carbon monomers and dimers into the liquid Cu surface, growth proceeds predominantly via the attachment of the former species. Dimer attachment as a possible source of defect formation at solid Cu is thus already reduced, and a self-healing mechanism of formed five-membered rings could further reduce defects at the graphene edges on the liquid surface.

These findings thus profoundly advance our comprehension of the atomistic processes involved in the CVD growth of graphene on a liquid copper surface. This enhanced understanding holds substantial significance for the ongoing development of 2D materials synthesis technologies.

METHODS

**Experimental details.** We used a customized CVD reactor capable of multi-technique *in situ* monitoring to investigate the graphene growth on a liquid copper catalyst under CVD conditions.[36] As the substrate, we used copper foils of high purity (99.9976%) purchased from Advent Research Materials (Eynsham, The United Kingdom) and tungsten disks from Metel BV (Waalwijk, The Netherlands) to support the molten copper. Before the first growth, we conditioned the copper foils by melting and etching them in a mixture of gaseous $H_2$ (9%) and Ar (91%) at a temperature $T \approx 1370$ K for a few hours to remove oxides and bulk impurities. The exact gas flows were controlled using Bronkhorst mass flow controllers and a residual gas analyzer (RGA). The gas partial pressures were calculated based on the gas correction factors (GCF) and the known total pressure in the reactor. Argon and hydrogen were constantly flown during operation with flows of 200 and 20 sccm, respectively. The total pressure in the reactor was kept at 200 mbar. We then proceeded with the growth of graphene using a 2% gas mixture of methane in argon as the gas precursor. We varied its flow between 0 and 26 sccm, corresponding to partial pressure ratios $p_{CH4}/p_{H2}$ between 0 and $2.72 \times 10^{-2}$. The graphene was grown on molten copper at the following temperatures $T$: 1368, 1399, 1416, 1433, and 1456 K, with an uncertainty of 5 K. At higher $CH_4$ flows, growth occurs too rapidly to be thoroughly analyzed. Nevertheless, we extended the experimental range of $p_{CH4}/p_{H2}$ by using a 5% methane concentration in argon to probe the range with the prevailing methane pressure based on the time required to cover the surface.

We monitored the CVD growth of graphene flakes on the liquid copper surface in real-time with a digital optical microscope used in radiation mode mounted above a quartz window of the reactor.[14] We recorded the microscopic images using a CMOS-based digital camera (frame rate of 0.5 Hz) and analyzed them using scripts written in MATLAB software.

**Computational details.** Molecular simulations were performed *via* a moment tensor potential (MTP)[62,63] for the Cu-C system, which is trained to the density functional theory (DFT) data computed with the Perdew-Burke-Ernzerhof (PBE) exchange-correlation functional[64] and the many-body dispersion (MBD) correction (PBE+MBD)[65]. This combination of machine learning potential and DFT has been demonstrated to be accurate and efficient in our previous work.[15] To describe more complicated configurations encountered in the studied chemical reactions, we extended our previous potential by an active learning framework based on furthest point sampling as described in detail in the SI.[66]

Using the trained potential combined with the umbrella sampling approach, we simulated free-energy surfaces of three crucial processes during graphene growth at the liquid copper surface: the decomposition and formation of one carbon dimer from/to two monomers and the attachment of a carbon monomer or a dimer to graphene zigzag and armchair edges. As a collective variable (CV), we use the minimum distance between carbon species and the graphene ribbon for the attachment processes and the monomer distance for the dimer dissociation. For each free-energy surface, the CV space is sliced into multiple narrow windows, and a biased simulation of 2 ns is performed in the canonical (NVT) ensemble at 1370 K in each window. We devise a simple parametric mean-field microkinetic model from the computed barriers to evaluate the kinetic competition between monomer attachment, dimer formation, and subsequent attachment. For more details and validation of the umbrella sampling simulations and the microkinetic model, see the SI.

ASSOCIATED CONTENT

**Supporting Information**. The Supporting Information is available free of charge. Movie S1 illustrates a typical growth procedure with varying $CH_4/H_2$ ratios, as presented in Figure 2a−d (AVI).

Movie S2 illustrates later growth stages with flake coalescence, as presented in Figure S9b−e (AVI).

Notes on CVD process, role of hydrogen, quality control by Raman spectroscopy, evolution of flake circumference and circularity, evolution of the flake size with temperature and gas flow, energy of etching, density functional theory calculations, training of machine learning potentials, free energy simulations, validation of minimum, microkinetic model of competing carbon monomer and dimer attachment (PDF).

AUTHOR INFORMATION

**Corresponding Author**

Valentina Rein − ESRF-The European Synchrotron, 38043 Grenoble, France; orcid.org/0000-0002-8142-2090; Email: valentina.belova@esrf.fr

Hendrik H. Heenen − Fritz-Haber-Institut der Max-Planck-Gesellschaft, 14195 Berlin, Germany; Email: heenen@fhi.mpg.de

**Author Contributions**

The manuscript was written through contributions of all authors. All authors have given approval to the final version of the manuscript. ‡V. R., H. G., and H. H. contributed equally to this work.

**Funding Sources**

The following funding is acknowledged: European Union's Horizon 2020 research and innovation program under Grant Agreement No. 951943 (DirectSepa).

**Notes**

The authors declare no competing financial interest.

ACKNOWLEDGMENT

The authors thank the Horizon 2020 research and innovation program of the European Union under Grant Agreement No. 951943 (DirectSepa) for funding. The authors thank Dr. Elias Diesen for the help in free energy calculations.